\documentclass[pre,twocolumn,groupedaddress,showpacs,floatfix]{revtex4}

\usepackage[usenames, dvipsnames]{color}
\usepackage{graphicx}
\usepackage{bm}
\usepackage{amsmath}
\usepackage{amsfonts}
\usepackage{amssymb}
\usepackage{latexsym}

\usepackage{epsf} 
\usepackage[latin1]{inputenc}

\newlength{\bb}
\begin{document}

\title{The Regularizing Capacity of Metabolic Networks}

\author{Carsten \surname{Marr}}
\email{marr@bio.tu-darmstadt.de}
\author{Mark \surname{M\"uller-Linow}}
\affiliation{Bioinformatics Group, Department of Biology,
  Darmstadt University of Technology, D-64287 Darmstadt, Germany}
\author{Marc-Thorsten \surname{H\"utt} }
\affiliation{Computational Systems Biology, School of Engineering
  and Science, International University Bremen\footnote{Jacobs
    University Bremen as of spring 2007}, D-28759 Bremen,
Germany}

%\date{\today}

\begin{abstract}
Despite their topological complexity almost all functional properties
of metabolic networks can be derived from steady-state
dynamics. Indeed, many theoretical investigations (like flux-balance
analysis) rely on extracting function from steady states. This leads
to the interesting question, how metabolic networks avoid complex
dynamics and maintain a steady-state behavior.  Here, we expose
metabolic network topologies to binary dynamics generated by simple
local rules. We find that the networks' response is highly specific:
Complex dynamics are systematically reduced on metabolic networks
compared to randomized networks with identical degree sequences. Already small
topological modifications substantially enhance the capacity of a
network to host complex dynamic behavior and thus reduce its
regularizing potential. This exceptionally pronounced regularization
of dynamics encoded in the topology may explain, why steady-state
behavior is ubiquitous in metabolism.
\end{abstract}

\pacs{89.75.Kd, 82.39.-k, 05.45.-a}

\maketitle

%%%%%%%%%%%%%%%%%%%%%%%%%%%%%%%%%%%%%%%%
\section{Introduction}
The general notion of network biology \citep{barabasi04} proposes an
abstract view on biological systems: The components' complex
interaction pattern is represented by a mathematical graph consisting
of nodes and links. The aim is to understand universal features and
design principles of complex biological networks at a system-wide
level \citep{alon03,barabasi04,strogatz01}. Recent findings include
the ubiquity of heavy-tail degree distributions \citep{barabasi04}, a
'bow-tie' structure of certain network types \citep{ma03b,csete04},
the presence of modules \citep{ravasz02, tanaka05, guimera05} and a
similarity in motif content of functionally related networks
\citep{milo04}.  An important focus of research is to develop models
of graph construction, which yield similar statistical properties as
the real graphs \citep{barabasi99,carlson00,arita04,tanaka05}. This
modeling task is complicated by the fact that dynamic performance is a
criterion in the evolutionary shaping of some types of biological
networks \citep{klemm05,kashtan05}. Consequently, a current challenge
is to incorporate dynamics into this general framework, i.e.~to link
topology and dynamic function. For gene regulatory networks huge
progress has been made in the last few years in that regard: The motif
content of eukaryotic genetic networks \citep{milo04} has been shown
\citep{klemm05} to correlate with the dynamic robustness profile of
these motifs; the dominant branch of the largest attractor of a
reduced yeast cell cycle network under the framework of Boolean
dynamics coincides with the experimentally observed sequence of cell
states \citep{li04}. \newline
In the case of metabolic networks the situation is far more
involved. Since the dynamics of the metabolic concentrations lack an
approximately binary behavior, refined kinetic models have been
developed \citep[see, e.g.,][]{reder88,klipp05}. Due to the limited
knowledge of kinetic parameters, such models are often restricted to
sub-networks, lacking the large-scale perspective, unless strong
simplifications and abstractions are introduced in the dynamics. Many
approaches exploit an intriguing feature of metabolic network
dynamics: the convergence to a steady state. Flux balance analysis
(FBA) \citep{kauffman03fba,palsson06}, for instance, retains the
stoichiometric matrix (and, therefore, the interaction pattern of the
network) to formulate hypotheses on the overall performance of the
system, together with an objective function incorporating constraints
on the metabolic dynamics. This method and related steady-state
approaches have been successfully applied to the phenotypic prediction
of various wild-type species and mutants under different environmental
conditions \citep[see, e.g.,][]{edwards00,stelling02,famili03}. \\
Here, we address a fundamental topological question: How well is a
particular network designed to suppress complex dynamics? We select
rules, which lead to transient (i.e., non steady-state) binary
dynamics on metabolic networks and compare the pattern
complexity for the real and modified topologies. Note that obviously,
our dynamics has no immediate connection to real metabolic dynamics,
it rather serves as a dynamic probe on complex networks. We find that
-- in spite of the structural and dynamical abstractions -- real
metabolic network topologies exhibit the capacity to maximally
regularize an imposed complex dynamics compared randomized
topologies with identical degree sequences.
\begin{figure*}[tbp]
\begin{center}
\includegraphics[width=16cm]{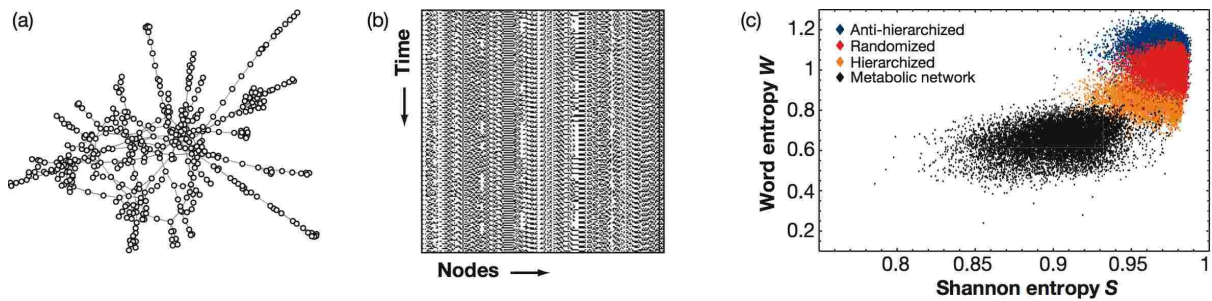}
\caption{(a) A graph representation of the largest connected component
  of the unipartite, substrate-centric metabolic network of yeast with
  $N=448$ nodes and $L=564$ links. (b) The spatio-temporal pattern for
  the $\Omega(\kappa_0)$ dynamics on the yeast metabolic network after
  a transient of $N$ time steps has been dropped. For visual clarity,
  we only show 200 nodes for 200 time-steps. (c) Entropy signature of
  the yeast metabolic network (black) and randomized (red),
  hierarchized (orange) and anti-hierarchized (blue) counterparts
  \citep[see][for details on these randomization
    procedures]{trusina04} with degree correlations of $0.05$, $0$,
  $+0.3$, and $-0.3$, respectively. Each point represents the entropy
  signature of the respective network for a randomly selected initial
  condition.}
\label{scheme}
\end{center}
\end{figure*}

%%%%%%%%%%%%%%%%%%%%%%%%%%%%%%%%%%%%%%%%
\section{Materials and Methods}
We use the metabolic networks compiled by Ma and Zeng (MZ)
\citep{ma03}. Their data rely on the KEGG database \citep{goto98} and
incorporate enzyme-catalyzed reactions based on genomic analyses and
biochemical literature. In this unipartite, substrate-centric
representation, metabolites (nodes) are connected by biochemical
reactions (links) whenever the catalyzing enzyme is encoded in the
respective genome. Compared to other metabolic network representations
the MZ networks are relatively sparse. This is due to the exclusion of
frequently occuring metabolites like ATP or NADH from those reactions,
where they act as current metabolites (or carrier metabolites;
cf. \cite{tanaka05}) only \citep[see][for a detailed discussion of
  current metabolites in metabolic networks]{ma03}. As noted in
\cite{ma03}, the exclusion of current metabolites is reasonable in
terms of metabolic pathways -- otherwise, for example, the path length
(number of reaction steps) from glucose to pyruvate in the glycolysis
pathway would reduce from nine to two \cite{ma03}. Since we are less
interested in mass flow and catalytic regulation, and more interested
in the information processing capabilities of such pathways, the
substrate-centric representation of biochemically meaningful pathways
of the MZ networks is an appropriate choice. However, we performed
parts of our analysis on other network representations (see Results)
with similar findings. From the available data we only use the largest
connected component of the networks similarly to previous studies
\citep{jeong00,ma03b,guimera05}. Hence, we limit our attention to the
most prominent part of the metabolic networks (on average, the largest
connected component comprises 54\% of the nodes and 66\% of the links
of the complete network for the 107 species in the MZ database) and
avoid dynamical artifacts due to isolated residual subgraphs of
different size.

%CA
Cellular automata (CA) are a well-known tool from complexity
theory. Under rather general assumptions symmetry arguments reduce the rule space for binary
cellular automata on graphs to a set of few outer-totalistic
\cite{packard85} automata. The effect of these assumptions (like
isotropy, locality and linearity) on the range of possible CA rules
will be discussed elsewhere \cite{inPrep}). Most of the rules lead to
trivial spatio-temporal patterns, e.g.~ oscillations or a steady state like the
frequently used majority rule \cite{moreira04,nochomovitz06}. Here, we
choose one of those rules, which leads to non steady-state behavior on
metabolic network topologies. Each node acts as a threshold device: the
state $x_i$ of node $i$ changes from 0 to 1 and vice versa, as soon as
the density of 1's in its neighborhood exceeds $\kappa$. $\Omega$
can be formalized to
\begin{equation}
  \label{omega}
  \Omega(\kappa): \; x_i(t+1) =
  \begin{cases}
    x_i(t) \,,    &  \rho_i \leq \kappa \\[0.2cm]
    1-x_i(t) \,,  &  \rho_i > \kappa \;.
  \end{cases}
\end{equation}  
The local density $\rho_i$ can be expressed by $\rho_i = \frac{1}{d_i}
\sum_j A_{ij}x_j$, where $d_i$ is the number of neighbors (the degree)
of node $i$ and $A$ denotes the adjacency matrix of the graph. The
rule $\Omega$ has been used previously to assess information
processing capacities of synthetic graphs \citep{marr05}. Complex
dynamics on MZ networks are achieved for a $\kappa_0$ around 0.3. The
exact size of the feasible interval depends on the maximal degree in
the network $d_{max}$, via $1/d_{max}$. Our implemented
complex dynamics can be considered as the continuous processing of
perturbations by the network. Following the argument of
\citep{wagner01} that perturbations can travel in both directions of
an irreversible reaction, we use undirected graphs as metabolic
network representations.

Information theory provides tools to analyze spatio-temporal patterns,
e.g., the Hamming distance, the mutual information or the Rényi
entropy \citep{infotheory}. Here we apply the Shannon entropy
\citep{shannon48} and the word entropy \cite{marr05}, since their
combination was found to perform well in separating different dynamic
regimes \citep{marr05}. We thus characterize a graph's capacity to
process binary information with the entropy signature ($S$,$W$) on the
basis of the specific update scheme $\Omega$. Entropy values are
calculated by analyzing $N$ time-steps for each of the $N$ nodes after
a transient time of $9 N$ time-steps.
\newline In our analysis, we take the mean Shannon entropy $S$ of the
$N$ individual entropies $S_i$, calculated for each node's time series
separately, as a measure for the structure of the overall pattern,
\begin{equation} \label{se}
   S =  \frac{1}{N} \sum_{i=1}^N S_i = \frac{1}{N} \sum_{i=1}^N - (p^0_i
   \log_2 p^0_i + p^1_i \log_2 p^1_i ) \;.
\end{equation}
The probabilities $p^0_i$ and $p^1_i$ denote the ratios of 0's and 1's
in the time series of node $i$. Constant nodes yield $S_i = 0$, while
nodes with a homogeneous distribution of 0's and 1's, that is ever
flipping and irregular nodes, contribute maximally to $S$ with $S_i
\approx 1$. In the dynamical regimes regarded in this study, the
value of $S$ is a measure for the homogeneity of dynamical behaviors
on the level of single states: Large values of $S$ emerge for an overall
oscillatory or complex dynamics, while smaller values of $S$ indicate
the existence of constant time series within the patterns. \newline
The word entropy $W$ serves as a simple and easily applicable
complexity measure of the emerging patterns beyond single time
steps. To quantify the irregularity of the time series of a single
node, we count the number of constant words, i.e.~time blocks of constant cell
states, of length $l$. The probability $p_i^l$ is the number of
words of length $l$ divided by the number of all constant blocks in
the time series of node $i$. The maximal possible word length is
simply the length $t$ of the time series analyzed. The word entropy
$W$ of a pattern is the average over the individual time series' entropies
$W_i$ and is defined as
 \begin{equation} \label{we}
 W  = \frac{1}{N} \sum_{i=1}^N W_i = \frac{1}{N} \sum_{i=1}^N \left( -
\sum_{l=1}^{t} p_i^l \log_2 p_i^l \right) \; .
\end{equation}
Patterns with solely oscillatory or stationary behaviors result in
$W$=0 while an overall irregular or complex behavior results in high
$W$. 

We use the term entropy for our observables due to their formal
definition and due to the application of similar concepts in
information theory \cite{shannon48} and the theory of cellular
automata (the word entropy serves as a feasible simplification of the
'block entropy', introduced in \cite{wolfram83}).  Note, however, that
the two entropy measures defined in eqs. (\ref{se}) and (\ref{we}) may not
be interpreted in the standard thermodynamical way, since we average
over the individual entropies of an ensemble of coupled dynamical
entities. Note also, that the $W_i$ are not bound to the interval
$[0,1]$.
\begin{figure}[tbp]
\begin{center}
\includegraphics[width=\bb]{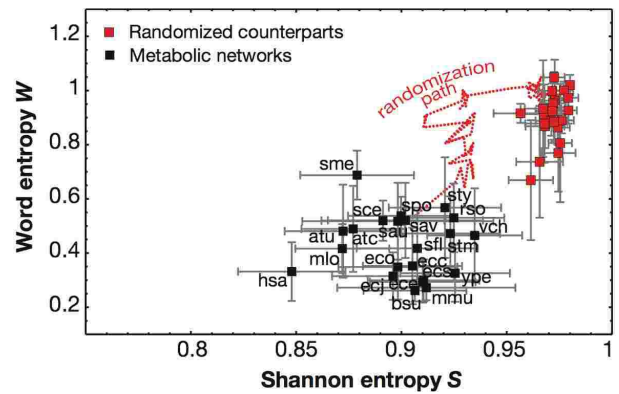}
\caption{Average entropy signature of the metabolic networks of 22
  species and randomized counterparts. The region, where the metabolic
  network topologies (black boxes) reside, is clearly separated from
  the entropy signatures of their randomized counterparts (red boxes)
  of same size, connectivity and degree sequence. $L$ randomization
  steps have been performed on every network. All shown networks have
  a link to node ratio $L/N \ge 1.22$. Error bars indicate the
  spreading of the entropy values due to different random initial
  conditions. Additional investigations showed that the separation in
  the plane is robust against variations of position and length of the
  analyzed pattern. A path between the entropy signature of the yeast
  and its randomized counterpart (dotted line) is obtained by stepwise
  randomization. The species abbreviations refer to the identifiers
  used in the MZ database.}
\label{selection}
\end{center}
\end{figure}

%%%%%%%%%%%%%%%%%%%%%%%%%%%%%%%%%%%%%%%%
\section{Results}
The metabolic network of the yeast, \textit{S. cerevisiae}, in the MZ
database comprises $N$=752 nodes and $L$=777 links. $N$=448 nodes and
$L$=564 links remain in the largest connected component of this
network.  It is characterized by a diverse mixture of linear chains of
nodes and hubs connecting chains and single elements (see
Fig.~\ref{scheme}a). This topological diversity is reproduced in the
patterns which emerge from running $\Omega(\kappa_0)$ on the network
(see Fig.~\ref{scheme}b): Constant and oscillatory time series coexist
with irregular time series, where nodes switch between alternating and
constant behavior in an unpredictable manner.  A randomized network,
where the topology is changed but the degree sequence of the graph is
preserved by always switching the links of two pairs of nodes as
proposed in \citep{maslov02}, shows patterns which are dominated by
irregularities. These differences in the dynamic response manifest
themselves in different entropy signatures of the yeast network and
randomized counterparts (see Fig.~\ref{scheme}c). The mean entropy
differences for $\Omega$ are: $\Delta S = S_{\textrm{rand}} -
S_{\textrm{yeast}} = 0.089$, $\Delta W = W_{\textrm{rand}} -
W_{\textrm{yeast}} = 0.50$. Apparently, the yeast topology is capable
of reducing the resulting entropies, that is, of regularizing the
dynamics imposed on it.  We checked that any rule among the cellular
automata on graphs set of rules \cite{inPrep} leading to complex
patterns on the yeast topology confirms the enhanced regularizing
capacity of the real metabolic network. Note that here and in the
following all graph modifications do not only conserve the overall
degree distribution but also the degree sequence, that is, the
individual degrees of all nodes. Moreover, we ensure that the
randomized graph is connected after every randomization step.

We probe the networks of all species in the MZ database and randomized
counterparts with the $\Omega(\kappa_0)$ dynamics. We find reduced
entropy signatures for the real topologies for 99 of the 107
species. The networks of the remaining 8 species are extremely sparse
and therefore difficult to properly randomize. A reduced $(S,W)$ is
still observed, if we exclude constant or trivially oscillating nodes
from the analysis. Not too sparse metabolic networks (we choose here a
link to node ratio $L/N \ge 1.22$) can be treated in a statistically
robust manner and cluster in the entropy plane at distinctly smaller
values than their randomized counterparts.\\
The modification of a network in single randomization steps but at
fixed initial conditions leads to a path within the entropy
plane. Consequently, this assessment of dynamic performance is highly
systematic: Small changes in graph topology lead to small changes in
the entropy signature.  In Fig.~\ref{selection}, such a randomization
procedure connects the metabolic network of the yeast and its
randomized counterpart. The opposite direction, from the randomized
counterpart towards the region of minimal entropies can in principle
be followed by an appropriate randomization process. In this sense,
the ($S$,$W$) entropy plane -- or an adequate set of other dynamic
observables -- may provide an orientation for simulated evolution or
purposive topological design.\\
\begin{figure}[tbp]
\begin{center}
\includegraphics[width=\bb]{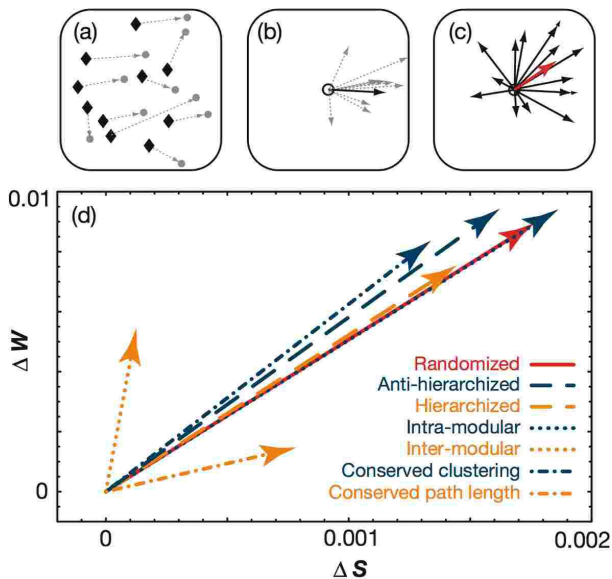}
\caption{Effect of minimal topological perturbations in yeast. Method:
  (a) For different random initial conditions, we yield entropy
  signatures for the original unperturbed network (black diamonds) and
  a minimally modified topology (gray dots); (b) The shifts for each
  initial condition (gray arrows) are averaged (black arrow); (c)
  Averaging over the ensemble of similarly modified topologies leads
  to the average entropy shift for this type of topological
  perturbation (red arrow). Result: (d) Systematic increase of the
  entropy shift with a single randomization step. Different protocols
  are performed, namely arbitrary randomization, randomization under
  conservation of clustering Coefficient and average path length
  respectively, hierarchization and anti-hierarchization,
  inter-modular and intra-modular randomization.}
\label{mutations}
\end{center}
\end{figure}

The randomization process alters a number of prominent topological
properties: The average path length and the clustering coefficient
\citep{watts98} are reduced, degree correlations \citep{trusina04} are
diminished and modular substructures \citep{ravasz02,guimera05} are
disintegrated. In order to investigate whether the optimization of the
metabolic networks goes beyond this set of topological observables, we
perform different minimal topological perturbations and assess the
corresponding entropy shift.
\newline We determine the entropy signatures of the yeast network and
a modified graph, where only a single modification step has been
performed, for different random initial conditions
(Fig.~\ref{mutations}a). Averaging over the initial conditions yields
the average entropy shift for one specific topological perturbation
(Fig.~\ref{mutations}b). Averaging over many topological perturbations
results in a vector which indicates the mean shift in the entropy
plane due to a specific modification protocol
(Fig.~\ref{mutations}c). We observe an average entropy increase for a
single randomization, hierarchization and anti-hierarchization step
(Fig.~\ref{mutations}d), where the latter two modifications, when
iterated, lead to positive and negative degree correlations,
respectively \citep[see][for an explanation of the corresponding
  algorithms]{trusina04}. In order to retain the modular structure of
the network we identify modules with a path length algorithm similar
to the one described in \citep{ma04} and randomize solely links within
or between single modules. Both modifications reveal an average
entropy increase compared to the original metabolic networks
(Fig.~\ref{mutations}d). In the case of intra-modular randomization,
the similarity to a random flip suggests a topological optimization
within the modules. The small shift due to inter-modular randomization
indicates an optimized structure on the highest modular scale. We
confirmed this result for another module identification algorithm
based on the topological overlap of nodes \citep{ravasz02}.  Moreover,
we confirmed an average entropy increase for protocols where a link
flip is only allowed if the average clustering coefficient is
conserved or increased due to a single randomization step
(Fig.~\ref{mutations}d). The same is true for a randomization protocol
which conserves the average path length
(Fig.~\ref{mutations}d). Again, we ensured that the randomized
networks remain connected through the application of the various
randomization protocols.  For all conditional randomization protocols,
a average entropy increase can result from a limited number of sampled
topologies. We checked, that this is not the case. Furthermore, the
entropy shift increases with the number of randomization steps,
i.e.~the modification depth, performed on the topology (data not
shown). Strongly hierarchized and anti-hierarchized networks (with
considerably positive and negative degree correlations of $+0.3$ and
$-0.3$ respectively), cluster in the entropy plane far apart from the
original networks and slightly below and above randomized networks
(see Fig.~\ref{scheme}c). \\

Summarizing, the different topological perturbations clearly show that
the pronounced regularizing capacity of real metabolic networks is not
trivially associated with a single topological property within the set
of graphs with a given degree distribution. Moreover, this property
goes beyond simple degree correlations and modularity and are
furthermore persistent on the level of individual modules. One can ask
if the reduced entropy signature of real metabolic networks compared
to randomized counterparts is an artifact of the specific network
representation we discuss throughout this article. However, we
verified the regularizing capacity of metabolic network topologies for
another uni-partite substrate-centric representation, where current
metabolites have not been removed from the system. Due to the
existence of these highly connected hubs, the 43 networks discussed in
\cite{jeong00} are highly connected and show extremely short average
path lengths. All systems with sufficient network size ($N>200$)
display a pronounced regularizing capacity compared to randomized
counterparts and an average entropy signature increase for singe
randomization steps. This is also true for two other networks we
explicitly checked: a directed version of the yeast network derived
from the MZ database and an undirected enzyme-centric network
based on the yeast stoichiometric matrix \cite{wwwpallson}, where two
enzymes are connected by a link if the product of the reaction
catalyzed by the first enzyme serves as an educt for the second.
Since a regularizing capacity is found for other representation of
metabolic networks as well, we think that this finding a generic
property of the network architecture of metabolic processes.

\begin{figure}[tbp]
\begin{center}
\includegraphics[width=\bb]{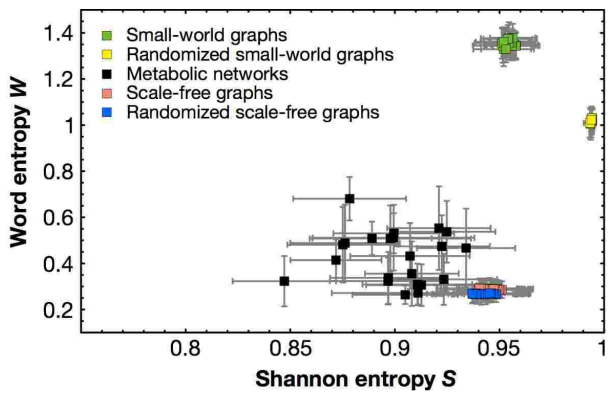}
\caption{Average entropy signature for the 22 species of
  Fig.~\ref{selection} together with synthetic architectures with the
  identical size and connectivity, respectively. We display data for
  the original metabolic networks (black), scale-free graphs (pink),
  randomized scale-free graphs (blue), small-world graphs (green) and
  randomized small-world graphs (yellow). The small-world graphs were
  generated by rewiring 5\% of the links of a regular chain. The
  randomized synthetic networks where obtained by flipping $L$ pairs
  of links. Notwithstanding the different network sizes and
  connectivities, the topological variants cluster in highly specific
  regions in the entropy plane. We averaged over 100 initial
  conditions for 100 different synthetic network realizations each. }
\label{synthetic}
\end{center}
\end{figure}
Passing from a real metabolic network to a randomized network samples
a subset of graphs with a particular degree sequence. Another level of
comparison is achieved if we regard the entropy signatures of simple
model graphs of the same network size and connectivity, that is, the
same number of nodes and links: Scale-free graphs obtained with a BA
scheme \cite{barabasi99} analog (where the number of attached links to
newly introduced node varies to achieve the desired number of links),
and regular graphs (where again the number of next-nearest neighbors
differs between two and three) after a few rewiring steps (small-world
graphs) according to the original Watts-Strogatz scheme
\cite{watts98}. Fig.~\ref{synthetic} shows the corresponding mean
entropy signature and standard deviation for the 22 species from
Fig.~\ref{selection}. Obviously, the synthetic model graphs react in a
highly specific manner to the imposed dynamics: They separate from the
metabolic networks and from each other and cluster tightly in the
entropy plane. The scale-free graphs exhibit a slightly reduced word
entropy compared to the real metabolic network, but an (on average)
higher Shannon entropy. For the small world graphs again both
entropies are elevated compared to the real metabolic
networks. Another important feature of Fig.~\ref{synthetic} is that
for the different network types the two entropies respond differently.
How do those model graphs react to randomization? For these synthetic
networks the entropy shift measures the bias of the dynamic response,
contained in the particular construction algorithm. For the scale-free
graphs, we find a small reduction of the average word entropy and
Shannon entropy under randomization. If we randomize small-world
graphs, and thereby break up the inherent regular neighborhoods, the
randomized networks exhibit a reduced average word entropy and a
elevated average Shannon entropy \cite{marr05}. Thus, an average
entropy increase, as observed for the metabolic networks discussed in
this paper, is not a trivial effect of the randomization procedure. On
the contrary, for chains of regularly connected nodes, a topological
perturbation in general leads to a decreasing dynamic complexity. In a
previous study \cite{marr06}, we showed this behaviour for synchronous
and asynchronous update schemes with different observables. Moreover
we showed, that both topological perturbation (i.e., the rewiring of
links) and dynamic noise increase the regularizing capacity of an
initially undisturbed cellular automaton.
%

%%%%%%%%%%%%%%%%%%%%%%%%%%%%%%%%%%%%%%%%
\section{Discussion}
Characterizing complex networks with dynamic probes is a novel and
promising approach.  The recent progress in understanding genetic
networks by mapping out the information flow through the corresponding
graphs constitutes an excellent example of this line of thought
\citep{bornholdt05}. Similarly, our analysis of metabolic networks
condenses a variety of topological features into the entropy signature
of the network, while the topological impact entering this quantity is
selected with respect to its dynamic effect. The method discussed here
may also serve as an unbiased probe to assess the dynamic performance
of different network types, like protein-interaction networks or
transcriptional networks, as well as other natural or technical
networks. Moreover, using system-specific tailored dynamics could
enhance the structure shaping process in engineering artificial
network topologies. \\
Our key result is that metabolic network topologies embody the
capacity to reliably regularize an imposed dynamics -- expressed by a
systematically reduced entropy signature compared to randomized
networks with identical degree sequences. The large-scale architecture
of metabolic networks is designed in such a way that complex and
chaotic dynamics are systematically dampened out.  We believe that the
architectural disposition for steady-state dynamics in metabolism
manifests in the observed regularizing capacity of the metabolic
network topologies.

In this investigation we focused on the entropies averaged over the
whole network. Particularly for a direct comparison with standard
analysis techniques of metabolic networks, we believe that the entropy
\textit{distribution} over the network could also be extremely
informative. The individual responses of the system's entities to a
dynamic probe might help characterize different contributions to the
functional state of the system, or, in terms of metabolic networks, it
might help understand the principles governing the reorganization of
steady-state fluxes under topological perturbations\cite{almaas05}.\\ \\

We acknowledge the comments of unknown referees, which substantially
improved the manuscript.

%%%%%%%%%%%%%%%%%%%%%%%%%%%%%%%%%%%%%%%%

%% \bibliography{/Users/carstenmarr/Documents/tex/bibtex/bib/literature}

\begin{thebibliography}{40}
\expandafter\ifx\csname natexlab\endcsname\relax\def\natexlab#1{#1}\fi
\expandafter\ifx\csname bibnamefont\endcsname\relax
  \def\bibnamefont#1{#1}\fi
\expandafter\ifx\csname bibfnamefont\endcsname\relax
  \def\bibfnamefont#1{#1}\fi
\expandafter\ifx\csname citenamefont\endcsname\relax
  \def\citenamefont#1{#1}\fi
\expandafter\ifx\csname url\endcsname\relax
  \def\url#1{\texttt{#1}}\fi
\expandafter\ifx\csname urlprefix\endcsname\relax\def\urlprefix{URL }\fi
\providecommand{\bibinfo}[2]{#2}
\providecommand{\eprint}[2][]{\url{#2}}

\bibitem[{\citenamefont{Barabási and Oltvai}(2004)}]{barabasi04}
\bibinfo{author}{\bibfnamefont{A.-L.} \bibnamefont{Barabási}} \bibnamefont{and}
  \bibinfo{author}{\bibfnamefont{Z.~N.} \bibnamefont{Oltvai}},
  \bibinfo{journal}{Nature Rev. Genet.} \textbf{\bibinfo{volume}{5}},
  \bibinfo{pages}{101} (\bibinfo{year}{2004}).

\bibitem[{\citenamefont{Alon}(2003)}]{alon03}
\bibinfo{author}{\bibfnamefont{U.}~\bibnamefont{Alon}},
  \bibinfo{journal}{Science} \textbf{\bibinfo{volume}{301}},
  \bibinfo{pages}{1866} (\bibinfo{year}{2003}).

\bibitem[{\citenamefont{Strogatz}(2001)}]{strogatz01}
\bibinfo{author}{\bibfnamefont{S.~H.} \bibnamefont{Strogatz}},
  \bibinfo{journal}{Nature} \textbf{\bibinfo{volume}{410}},
  \bibinfo{pages}{268} (\bibinfo{year}{2001}).

\bibitem[{\citenamefont{Ma and Zeng}(2003{\natexlab{a}})}]{ma03b}
\bibinfo{author}{\bibfnamefont{H.}~\bibnamefont{Ma}} \bibnamefont{and}
  \bibinfo{author}{\bibfnamefont{A.-P.} \bibnamefont{Zeng}},
  \bibinfo{journal}{Bioinformatics} \textbf{\bibinfo{volume}{19}},
  \bibinfo{pages}{1423} (\bibinfo{year}{2003}{\natexlab{a}}).

\bibitem[{\citenamefont{Csete and Doyle}(2004)}]{csete04}
\bibinfo{author}{\bibfnamefont{M.}~\bibnamefont{Csete}} \bibnamefont{and}
  \bibinfo{author}{\bibfnamefont{J.}~\bibnamefont{Doyle}},
  \bibinfo{journal}{Trends Biotech.} \textbf{\bibinfo{volume}{22}},
  \bibinfo{pages}{446} (\bibinfo{year}{2004}).

\bibitem[{\citenamefont{Ravasz et~al.}(2002)\citenamefont{Ravasz, Somera,
  Mongru, Oltvai, and Barabási}}]{ravasz02}
\bibinfo{author}{\bibfnamefont{E.}~\bibnamefont{Ravasz}},
  \bibinfo{author}{\bibfnamefont{A.~L.} \bibnamefont{Somera}},
  \bibinfo{author}{\bibfnamefont{D.~A.} \bibnamefont{Mongru}},
  \bibinfo{author}{\bibfnamefont{Z.~N.} \bibnamefont{Oltvai}},
  \bibnamefont{and} \bibinfo{author}{\bibfnamefont{A.-L.}
  \bibnamefont{Barabási}}, \bibinfo{journal}{Science}
  \textbf{\bibinfo{volume}{297}}, \bibinfo{pages}{1551} (\bibinfo{year}{2002}).

\bibitem[{\citenamefont{Tanaka}(2005)}]{tanaka05}
\bibinfo{author}{\bibfnamefont{R.}~\bibnamefont{Tanaka}},
  \bibinfo{journal}{Phys. Rev. Lett.} \textbf{\bibinfo{volume}{94}},
  \bibinfo{eid}{168101} (\bibinfo{year}{2005}).

\bibitem[{\citenamefont{Guimerà and Amaral}(2005)}]{guimera05}
\bibinfo{author}{\bibfnamefont{R.}~\bibnamefont{Guimerà}} \bibnamefont{and}
  \bibinfo{author}{\bibfnamefont{L.~A.~N.} \bibnamefont{Amaral}},
  \bibinfo{journal}{Nature} \textbf{\bibinfo{volume}{433}},
  \bibinfo{pages}{895} (\bibinfo{year}{2005}).

\bibitem[{\citenamefont{Milo et~al.}(2004)\citenamefont{Milo, Itzkovitz,
  Kashtan, Levitt, Shen-Orr, Ayzenshtat, Sheffer, and Alon}}]{milo04}
\bibinfo{author}{\bibfnamefont{R.}~\bibnamefont{Milo}},
  \bibinfo{author}{\bibfnamefont{S.}~\bibnamefont{Itzkovitz}},
  \bibinfo{author}{\bibfnamefont{N.}~\bibnamefont{Kashtan}},
  \bibinfo{author}{\bibfnamefont{R.}~\bibnamefont{Levitt}},
  \bibinfo{author}{\bibfnamefont{S.}~\bibnamefont{Shen-Orr}},
  \bibinfo{author}{\bibfnamefont{I.}~\bibnamefont{Ayzenshtat}},
  \bibinfo{author}{\bibfnamefont{M.}~\bibnamefont{Sheffer}}, \bibnamefont{and}
  \bibinfo{author}{\bibfnamefont{U.}~\bibnamefont{Alon}},
  \bibinfo{journal}{Science} \textbf{\bibinfo{volume}{303}},
  \bibinfo{pages}{1538} (\bibinfo{year}{2004}).

\bibitem[{\citenamefont{Barabási and Albert}(1999)}]{barabasi99}
\bibinfo{author}{\bibfnamefont{A.-L.} \bibnamefont{Barabási}} \bibnamefont{and}
  \bibinfo{author}{\bibfnamefont{R.}~\bibnamefont{Albert}},
  \bibinfo{journal}{Science} \textbf{\bibinfo{volume}{286}},
  \bibinfo{pages}{509} (\bibinfo{year}{1999}).

\bibitem[{\citenamefont{Carlson and Doyle}(2000)}]{carlson00}
\bibinfo{author}{\bibfnamefont{J.~M.} \bibnamefont{Carlson}} \bibnamefont{and}
  \bibinfo{author}{\bibfnamefont{J.}~\bibnamefont{Doyle}},
  \bibinfo{journal}{Phys. Rev. Lett.} \textbf{\bibinfo{volume}{84}},
  \bibinfo{pages}{2529} (\bibinfo{year}{2000}).

\bibitem[{\citenamefont{Arita}(2004)}]{arita04}
\bibinfo{author}{\bibfnamefont{M.}~\bibnamefont{Arita}},
  \bibinfo{journal}{Proc. Natl. Acad. Sci. USA} \textbf{\bibinfo{volume}{101}},
  \bibinfo{pages}{1543} (\bibinfo{year}{2004}).

\bibitem[{\citenamefont{Klemm and Bornholdt}(2005)}]{klemm05}
\bibinfo{author}{\bibfnamefont{K.}~\bibnamefont{Klemm}} \bibnamefont{and}
  \bibinfo{author}{\bibfnamefont{S.}~\bibnamefont{Bornholdt}},
  \bibinfo{journal}{Proc. Natl. Acad. Sci. USA} \textbf{\bibinfo{volume}{102}},
  \bibinfo{pages}{18414} (\bibinfo{year}{2005}).

\bibitem[{\citenamefont{Kashtan and Alon}(2005)}]{kashtan05}
\bibinfo{author}{\bibfnamefont{N.}~\bibnamefont{Kashtan}} \bibnamefont{and}
  \bibinfo{author}{\bibfnamefont{U.}~\bibnamefont{Alon}},
  \bibinfo{journal}{Proc. Natl. Acad. Sci. USA} \textbf{\bibinfo{volume}{102}},
  \bibinfo{pages}{13773} (\bibinfo{year}{2005}).

\bibitem[{\citenamefont{Li et~al.}(2004)\citenamefont{Li, Long, Lu, Ouyang, and
  Tang}}]{li04}
\bibinfo{author}{\bibfnamefont{F.}~\bibnamefont{Li}},
  \bibinfo{author}{\bibfnamefont{T.}~\bibnamefont{Long}},
  \bibinfo{author}{\bibfnamefont{Y.}~\bibnamefont{Lu}},
  \bibinfo{author}{\bibfnamefont{Q.}~\bibnamefont{Ouyang}}, \bibnamefont{and}
  \bibinfo{author}{\bibfnamefont{C.}~\bibnamefont{Tang}},
  \bibinfo{journal}{Proc. Natl. Acad. Sci. USA} \textbf{\bibinfo{volume}{101}},
  \bibinfo{pages}{4781} (\bibinfo{year}{2004}).

\bibitem[{\citenamefont{Reder}(1988)}]{reder88}
\bibinfo{author}{\bibfnamefont{C.}~\bibnamefont{Reder}}, \bibinfo{journal}{J.
  Theor. Biol.} \textbf{\bibinfo{volume}{135}}, \bibinfo{pages}{175}
  (\bibinfo{year}{1988}).

\bibitem[{\citenamefont{Klipp et~al.}(2005)\citenamefont{Klipp, Herwig, Kowald,
  Wierling, and Lehrach}}]{klipp05}
\bibinfo{author}{\bibfnamefont{E.}~\bibnamefont{Klipp}},
  \bibinfo{author}{\bibfnamefont{R.}~\bibnamefont{Herwig}},
  \bibinfo{author}{\bibfnamefont{A.}~\bibnamefont{Kowald}},
  \bibinfo{author}{\bibfnamefont{C.}~\bibnamefont{Wierling}}, \bibnamefont{and}
  \bibinfo{author}{\bibfnamefont{H.}~\bibnamefont{Lehrach}},
  \emph{\bibinfo{title}{{Systems Biology in Practice}}}
  (\bibinfo{publisher}{Wiley}, \bibinfo{address}{Weinheim},
  \bibinfo{year}{2005}).

\bibitem[{\citenamefont{Kauffman et~al.}(2003)\citenamefont{Kauffman, Prakash,
  and Edwards}}]{kauffman03fba}
\bibinfo{author}{\bibfnamefont{K.~J.} \bibnamefont{Kauffman}},
  \bibinfo{author}{\bibfnamefont{P.}~\bibnamefont{Prakash}}, \bibnamefont{and}
  \bibinfo{author}{\bibfnamefont{J.~S.} \bibnamefont{Edwards}},
  \bibinfo{journal}{Curr. Opin. Biotech.} \textbf{\bibinfo{volume}{14}},
  \bibinfo{pages}{491} (\bibinfo{year}{2003}).

\bibitem[{\citenamefont{Palsson}(2006)}]{palsson06}
\bibinfo{author}{\bibfnamefont{B.}~\bibnamefont{Palsson}},
  \emph{\bibinfo{title}{{Systems biology: properties of reconstructed
  networks}}} (\bibinfo{publisher}{Cambridge University Press},
  \bibinfo{address}{New York}, \bibinfo{year}{2006}).

\bibitem[{\citenamefont{Edwards and Palsson}(2000)}]{edwards00}
\bibinfo{author}{\bibfnamefont{J.~S.} \bibnamefont{Edwards}} \bibnamefont{and}
  \bibinfo{author}{\bibfnamefont{B.~O.} \bibnamefont{Palsson}},
  \bibinfo{journal}{Proc. Natl. Acad. Sci. USA} \textbf{\bibinfo{volume}{97}},
  \bibinfo{pages}{5528} (\bibinfo{year}{2000}).

\bibitem[{\citenamefont{Stelling et~al.}(2002)\citenamefont{Stelling, Klamt,
  Bettenbrock, Schuster, and Gilles}}]{stelling02}
\bibinfo{author}{\bibfnamefont{J.}~\bibnamefont{Stelling}},
  \bibinfo{author}{\bibfnamefont{S.}~\bibnamefont{Klamt}},
  \bibinfo{author}{\bibfnamefont{K.}~\bibnamefont{Bettenbrock}},
  \bibinfo{author}{\bibfnamefont{S.}~\bibnamefont{Schuster}}, \bibnamefont{and}
  \bibinfo{author}{\bibfnamefont{E.}~\bibnamefont{Gilles}},
  \bibinfo{journal}{Nature} \textbf{\bibinfo{volume}{420}},
  \bibinfo{pages}{190} (\bibinfo{year}{2002}).

\bibitem[{\citenamefont{Famili et~al.}(2003)\citenamefont{Famili, Förster,
  Nielsen, and Palsson}}]{famili03}
\bibinfo{author}{\bibfnamefont{I.}~\bibnamefont{Famili}},
  \bibinfo{author}{\bibfnamefont{J.}~\bibnamefont{Förster}},
  \bibinfo{author}{\bibfnamefont{J.}~\bibnamefont{Nielsen}}, \bibnamefont{and}
  \bibinfo{author}{\bibfnamefont{B.}~\bibnamefont{Palsson}},
  \bibinfo{journal}{Proc. Natl. Acad. Sci. USA} \textbf{\bibinfo{volume}{100}},
  \bibinfo{pages}{13134} (\bibinfo{year}{2003}).

\bibitem[{\citenamefont{Trusina et~al.}(2004)\citenamefont{Trusina, Maslov,
  Minnhagen, and Sneppen}}]{trusina04}
\bibinfo{author}{\bibfnamefont{A.}~\bibnamefont{Trusina}},
  \bibinfo{author}{\bibfnamefont{S.}~\bibnamefont{Maslov}},
  \bibinfo{author}{\bibfnamefont{P.}~\bibnamefont{Minnhagen}},
  \bibnamefont{and} \bibinfo{author}{\bibfnamefont{K.}~\bibnamefont{Sneppen}},
  \bibinfo{journal}{Phys. Rev. Lett.} \textbf{\bibinfo{volume}{92}},
  \bibinfo{pages}{178702} (\bibinfo{year}{2004}).

\bibitem[{\citenamefont{Ma and Zeng}(2003{\natexlab{b}})}]{ma03}
\bibinfo{author}{\bibfnamefont{H.}~\bibnamefont{Ma}} \bibnamefont{and}
  \bibinfo{author}{\bibfnamefont{A.-P.} \bibnamefont{Zeng}},
  \bibinfo{journal}{Bioinformatics} \textbf{\bibinfo{volume}{19}},
  \bibinfo{pages}{270} (\bibinfo{year}{2003}{\natexlab{b}}).

\bibitem[{\citenamefont{Goto et~al.}(1998)\citenamefont{Goto, Nishioka, and
  Kanehisa}}]{goto98}
\bibinfo{author}{\bibfnamefont{S.}~\bibnamefont{Goto}},
  \bibinfo{author}{\bibfnamefont{T.}~\bibnamefont{Nishioka}}, \bibnamefont{and}
  \bibinfo{author}{\bibfnamefont{M.}~\bibnamefont{Kanehisa}},
  \bibinfo{journal}{Bioinformatics} \textbf{\bibinfo{volume}{14}},
  \bibinfo{pages}{591} (\bibinfo{year}{1998}).

\bibitem[{\citenamefont{Jeong et~al.}(2000)\citenamefont{Jeong, Tombor, Albert,
  Oltvai, and Barabási}}]{jeong00}
\bibinfo{author}{\bibfnamefont{H.}~\bibnamefont{Jeong}},
  \bibinfo{author}{\bibfnamefont{B.}~\bibnamefont{Tombor}},
  \bibinfo{author}{\bibfnamefont{R.}~\bibnamefont{Albert}},
  \bibinfo{author}{\bibfnamefont{Z.~N.} \bibnamefont{Oltvai}},
  \bibnamefont{and} \bibinfo{author}{\bibfnamefont{A.-L.}
  \bibnamefont{Barabási}}, \bibinfo{journal}{Nature}
  \textbf{\bibinfo{volume}{407}}, \bibinfo{pages}{651} (\bibinfo{year}{2000}).

\bibitem[{\citenamefont{Packard and Wolfram}(1985)}]{packard85}
\bibinfo{author}{\bibfnamefont{N.~H.} \bibnamefont{Packard}} \bibnamefont{and}
  \bibinfo{author}{\bibfnamefont{S.}~\bibnamefont{Wolfram}},
  \bibinfo{journal}{Journal of Statistical Physics}
  \textbf{\bibinfo{volume}{38}}, \bibinfo{pages}{901} (\bibinfo{year}{1985}).

\bibitem[{\citenamefont{Marr and H\"utt}(2007)}]{inPrep}
\bibinfo{author}{\bibfnamefont{C.} \bibnamefont{Marr}} \bibnamefont{and}
  \bibinfo{author}{\bibfnamefont{M.-Th.}~\bibnamefont{H\"utt}} (in preparation).

\bibitem[{\citenamefont{Moreira et~al.}(2004)\citenamefont{Moreira, Mathur, and
  Amaral}}]{moreira04}
\bibinfo{author}{\bibfnamefont{A.~A.} \bibnamefont{Moreira}},
  \bibinfo{author}{\bibfnamefont{A.}~\bibnamefont{Mathur}}, \bibnamefont{and}
  \bibinfo{author}{\bibfnamefont{D.~D. L.~A.~N.} \bibnamefont{Amaral}},
  \bibinfo{journal}{Proc. Natl. Acad. Sci. USA} \textbf{\bibinfo{volume}{101}},
  \bibinfo{pages}{12085} (\bibinfo{year}{2004}).

\bibitem[{\citenamefont{Nochomovitz and Li}(2006)}]{nochomovitz06}
\bibinfo{author}{\bibfnamefont{Y.}~\bibnamefont{Nochomovitz}} \bibnamefont{and}
  \bibinfo{author}{\bibfnamefont{H.}~\bibnamefont{Li}}, \bibinfo{journal}{Proc.
  Natl. Acad. Sci. USA} \textbf{\bibinfo{volume}{103}}, \bibinfo{pages}{4180}
  (\bibinfo{year}{2006}).

\bibitem[{\citenamefont{Marr and Hütt}(2005)}]{marr05}
\bibinfo{author}{\bibfnamefont{C.}~\bibnamefont{Marr}} \bibnamefont{and}
  \bibinfo{author}{\bibfnamefont{M.-T.} \bibnamefont{Hütt}},
  \bibinfo{journal}{Physica A} \textbf{\bibinfo{volume}{354}},
  \bibinfo{pages}{641} (\bibinfo{year}{2005}).

\bibitem[{\citenamefont{Wagner and Fell}(2001)}]{wagner01}
\bibinfo{author}{\bibfnamefont{A.}~\bibnamefont{Wagner}} \bibnamefont{and}
  \bibinfo{author}{\bibfnamefont{D.~A.} \bibnamefont{Fell}},
  \bibinfo{journal}{Proc. R. Soc. B} \textbf{\bibinfo{volume}{268}},
  \bibinfo{pages}{1803} (\bibinfo{year}{2001}).

\bibitem[{\citenamefont{Jones and Jones}(2000)}]{infotheory}
\bibinfo{author}{\bibfnamefont{G.~A.} \bibnamefont{Jones}} \bibnamefont{and}
  \bibinfo{author}{\bibfnamefont{J.~M.} \bibnamefont{Jones}},
  \emph{\bibinfo{title}{Information and Coding Theory}}
  (\bibinfo{publisher}{Springer}, \bibinfo{address}{Berlin},
  \bibinfo{year}{2000}).

\bibitem[{\citenamefont{Shannon}(1948)}]{shannon48}
\bibinfo{author}{\bibfnamefont{C.~E.} \bibnamefont{Shannon}},
  \bibinfo{journal}{The Bell Systems Technical Journal}
  \textbf{\bibinfo{volume}{27}}, \bibinfo{pages}{379} (\bibinfo{year}{1948}).

\bibitem[{\citenamefont{Wolfram}(1983)}]{wolfram83}
\bibinfo{author}{\bibfnamefont{S.}~\bibnamefont{Wolfram}},
  \bibinfo{journal}{Rev. Mod. Phys.} \textbf{\bibinfo{volume}{55}},
  \bibinfo{pages}{601} (\bibinfo{year}{1983}).

\bibitem[{\citenamefont{Maslov and Sneppen}(2002)}]{maslov02}
\bibinfo{author}{\bibfnamefont{S.}~\bibnamefont{Maslov}} \bibnamefont{and}
  \bibinfo{author}{\bibfnamefont{K.}~\bibnamefont{Sneppen}},
  \bibinfo{journal}{Science} \textbf{\bibinfo{volume}{296}},
  \bibinfo{pages}{910} (\bibinfo{year}{2002}).

\bibitem[{\citenamefont{Watts and Strogatz}(1998)}]{watts98}
\bibinfo{author}{\bibfnamefont{D.~J.} \bibnamefont{Watts}} \bibnamefont{and}
  \bibinfo{author}{\bibfnamefont{S.~H.} \bibnamefont{Strogatz}},
  \bibinfo{journal}{Nature} \textbf{\bibinfo{volume}{393}},
  \bibinfo{pages}{440} (\bibinfo{year}{1998}).

\bibitem[{\citenamefont{Ma et~al.}(2004)\citenamefont{Ma, Zhao, Yuan, and
  Zeng}}]{ma04}
\bibinfo{author}{\bibfnamefont{H.~W.} \bibnamefont{Ma}},
  \bibinfo{author}{\bibfnamefont{X.~M.} \bibnamefont{Zhao}},
  \bibinfo{author}{\bibfnamefont{Y.~J.} \bibnamefont{Yuan}}, \bibnamefont{and}
  \bibinfo{author}{\bibfnamefont{A.~P.} \bibnamefont{Zeng}},
  \bibinfo{journal}{Bioinformatics} \textbf{\bibinfo{volume}{20}},
  \bibinfo{pages}{1870} (\bibinfo{year}{2004}).

\bibitem[{\citenamefont{pallson www}(2006)}]{wwwpallson}
\bibinfo{note} The stoichiometric matrices of various organisms
including yeast can be
downloaded in SMBL at http://gcrg.ucsd.edu/organisms/index.html.

%% \bibitem[{\citenamefont{Erd\H{o}s and Rényi}(1959)}]{erdos59}
%% \bibinfo{author}{\bibfnamefont{P.}~\bibnamefont{Erd\H{o}s}} \bibnamefont{and}
%%   \bibinfo{author}{\bibfnamefont{A.}~\bibnamefont{Rényi}},
%%   \bibinfo{journal}{Publicationes Mathematicae} \textbf{\bibinfo{volume}{6}},
%%   \bibinfo{pages}{290} (\bibinfo{year}{1959}).

\bibitem[{\citenamefont{Marr and Hütt}(2006)}]{marr06}
\bibinfo{author}{\bibfnamefont{C.}~\bibnamefont{Marr}} \bibnamefont{and}
  \bibinfo{author}{\bibfnamefont{M.-T.} \bibnamefont{Hütt}},
  \bibinfo{journal}{Phys. Lett. A} \textbf{\bibinfo{volume}{349}},
  \bibinfo{pages}{302} (\bibinfo{year}{2006}).

\bibitem[{\citenamefont{Bornholdt}(2005)}]{bornholdt05}
\bibinfo{author}{\bibfnamefont{S.}~\bibnamefont{Bornholdt}},
  \bibinfo{journal}{Science} \textbf{\bibinfo{volume}{310}},
  \bibinfo{pages}{449} (\bibinfo{year}{2005}).

\bibitem[{\citenamefont{Almaas et~al.}(2005)\citenamefont{Almaas, Oltvai, and
  Barabási}}]{almaas05}
\bibinfo{author}{\bibfnamefont{E.}~\bibnamefont{Almaas}},
  \bibinfo{author}{\bibfnamefont{Z.}~\bibnamefont{Oltvai}}, \bibnamefont{and}
  \bibinfo{author}{\bibfnamefont{A.-L.} \bibnamefont{Barabási}},
  \bibinfo{journal}{PLoS Comput. Biol.} \textbf{\bibinfo{volume}{1}},
  \bibinfo{pages}{e68} (\bibinfo{year}{2005}).

\end{thebibliography}
%% \end{document}

\end{document}